\documentstyle[12pt]{article}
\newcommand{\svec}[1]{ \stackrel{\rightarrow}{#1} }

\newcommand{\define}{ \stackrel{\triangle}{=} }

\def\be{\begin{equation}}
\def\ee{\end{equation}}
\def\ba{\begin{array}}
\def\ea{\end{array}}

\begin{document}
\title{\bf Classical Gravitational Interactions and
            Gravitational Lorentz Force }
\author{{Ning Wu}
\thanks{email address: wuning@mail.ihep.ac.cn}
\\
\\
{\small Institute of High Energy Physics, P.O.Box 918-1,
Beijing 100039, P.R.China}}
\maketitle
\vskip 0.8in

~~

\vskip 0.4in

\begin{abstract}
In quantum gauge theory of gravity, the gravitational field
is represented by gravitational gauge field. The field strength
of gravitational gauge field has both gravitoelectric
component and gravitomagnetic component. In classical
level, gauge theory of gravity gives out classical Newtonian
gravitational interactions in a relativistic form. Besides,
it gives out gravitational Lorentz force which is the gravitational
force on a moving object in gravitomagnetic field.
The direction of gravitational Lorentz force does not along
that of classical gravitational Newtonian force.
Effects of gravitational Lorentz force should be detectable,
and these effects can be used to discriminate gravitomagnetic
field from ordinary electromagnetic magnetic field.
\\

\end{abstract}

\newpage

\Roman{section}

\section{Introduction}
\setcounter{equation}{0}

In classical Newton's theory of gravity, gravity obeys the
inverse square law\cite{01}. That is, the gravitational force
between two mass point $m_1$ and $m_2$ is
\be
f= \frac{G_N m_1 m_2}{r^2},
\ee
where $G_N$ is the newtonian gravitational constant and $r$ is
the distance between two mass point. Gravitational force is along
the line which connects the mass point and is always attractive.
Obvious, the above expression is not invariant  or covariant
under Lorentz transformation. \\

In Einstein's general theory of relativity\cite{02,03}, gravity is treated
as space-time geometry, and gravity is just an effect of the
curved space-time. General relativity is a geometric theory
of gravity, and in the original expressions, the concepts of
the space-time metric, affine connection, curvature
tensor, geodesic curve, $\cdots$ are used, and the concept of
the traditional "gravitational force" is not used. Only in
post-Newtonian approximation can we clearly see the correspondent
of traditional gravitational force and its influences\cite{04,05}. \\

Quantum gauge theory of gravity is proposed in the framework of the
quantum field theory\cite{wu01,wu02,wu03,wu04,wu05},
where gravity is treated as a kind of physical interactions
and space-time is always kept flat. This treatment satisfies
the fundamental spirit of quantum field theory.
Quantum gauge theory of gravity is a perturbatively
renormalizable quantum theory. Quantum gauge theory can be
used to solve some problems related to quantum behavior of
gravitational interactions, such as unifications of fundamental
interactions can be easily accomplished in it\cite{12,13,14},
it can be used to explain the possible origin of dark matter
and dard energy\cite{15,16}, it can be used to explain
Podkletnov effects\cite{17} and COW experiments. \\

In gauge theory of gravity, gravitoelectromagnetic field
is naturally defined as components of field strength of
gravitational gauge field. It is known that gravitoelectromagnetism
was studied for more than 130 years.
The close analogy between Newton's gravitation law and Coulomb's law
of electricity led to the birth of the concept of gravitomagnetism
in the nineteenth century\cite{m01,m02,m03,m04}. Later, in the
framework of General Relativity, gravitomagnetism was extensively
explored\cite{m05,m06,m07} and recently some experiments are
designed to test gravitomagnetism effects. Some recently reviews
on gravitomagnetism can be found in literatures \cite{m08,m09,m10}.
In quantum gauge theory of gravity, gravitoelectromagnetism
is defined in a more general way, and the gravitoelectromagnetism
discussed in the literatures \cite{m05,m06,m07,m08,m09,m10} is
only the special case of it. \\

In this paper, we will discuss some classical effects of
gravitational interactions in the framework of gauge theory
of gravity, including gravitational interactions of two mass
point and gravitational Lorentz force. \\

\section{Pure Gravitational Gauge Field}
\setcounter{equation}{0}

First, for the sake of integrity, we give a simple introduction
to gravitational gauge theory and introduce some notations which
is used in this paper. Details on quantum gauge theory of gravity
can be found in literatures \cite{wu01,wu02,wu03,wu04,wu05}.
In gauge theory of gravity, the most fundamental
quantity is gravitational gauge field $C_{\mu}(x)$,
which is the gauge potential corresponding to gravitational
gauge symmetry. Gauge field $C_{\mu}(x)$ is a vector in
the corresponding Lie algebra, which
is called the gravitational Lie algebra.
So $C_{\mu}(x)$ can expanded as
\be \label{2.10}
C_{\mu}(x) = C_{\mu}^{\alpha} (x) \hat{P}_{\alpha},
~~~~~~(\mu, \alpha = 0,1,2,3)
\ee
where $C_{\mu}^{\alpha}(x)$ is the component field and
$\hat{P}_{\alpha} = -i \frac{\partial}{\partial x^{\alpha}}$
is the  generator of the global gravitational gauge group.
The gravitational gauge covariant derivative is defined by
\be \label{2.9}
D_{\mu} = \partial_{\mu} - i g C_{\mu} (x)
= G_{\mu}^{\alpha} \partial_{\alpha},
\ee
where $g$ is the gravitational coupling constant and
$G$ is given by
\be \label{2.11}
G = (G_{\mu}^{\alpha}) = ( \delta_{\mu}^{\alpha} - g C_{\mu}^{\alpha} )
=(I -g C)^{\alpha}_{\mu}.
\ee
Matrix $G$ is an important quantity in gauge theory
of gravity. Its inverse matrix is denoted as $G^{-1}$
\be \label{2.12}
G^{-1} = \frac{1}{I - g C} = (G^{-1 \mu}_{\alpha}).
\ee
Using matrix $G$ and $G^{-1}$, we can define two important
quantities
\be \label{2.13}
g^{\alpha \beta} = \eta^{\mu \nu}
G^{\alpha}_{\mu} G^{\beta}_{\nu},
\ee
\be \label{2.14}
g_{\alpha \beta} = \eta_{\mu \nu}
G_{\alpha}^{-1 \mu} G_{\beta}^{-1 \nu}.
\ee
\\

The  field strength of gravitational gauge field is defined by
\be \label{2.16}
F_{\mu\nu} \define \frac{1}{-ig} \lbrack D_{\mu}~~,~~D_{\nu} \rbrack.
\ee
Its explicit expression is
\be \label{2.17}
F_{\mu\nu}(x) = \partial_{\mu} C_{\nu} (x)
-\partial_{\nu} C_{\mu} (x)
- i g C_{\mu} (x) C_{\nu}(x)
+ i g C_{\nu} (x) C_{\mu}(x).
\ee
$F_{\mu\nu}$ is also a vector in gravitational Lie algebra
and can be expanded as,
\be \label{2.18}
F_{\mu\nu} (x) = F_{\mu\nu}^{\alpha}(x) \cdot \hat{P}_{\alpha},
\ee
where
\be \label{2.19}
F_{\mu\nu}^{\alpha} = \partial_{\mu} C_{\nu}^{\alpha}
-\partial_{\nu} C_{\mu}^{\alpha}
-  g C_{\mu}^{\beta} \partial_{\beta} C_{\nu}^{\alpha}
+  g C_{\nu}^{\beta} \partial_{\beta} C_{\mu}^{\alpha}.
\ee
\\

\section{Gravitoelectromagnetic Field}
\setcounter{equation}{0}

$F_{\mu \nu}^{\alpha}$ is the component field strength of
gravitational gauge field.  Define
\be
F^{\alpha}_{ij}= - \varepsilon_{ijk} B^{\alpha}_{k}
~~,~~
F^{\alpha}_{ 0i} = E^{\alpha}_{i}.
\label{12.2}
\ee
Then field strength of gravitational gauge field can be expressed
as
\be
F^{\alpha} = \left \lbrace
\ba{cccc}
0 & E_1^{\alpha} & E_2^{\alpha} & E_3^{\alpha}  \\
- E_1^{\alpha} & 0 & -B_3^{\alpha}  &  B_2^{\alpha}  \\
- E_2^{\alpha} & B_3^{\alpha} & 0  & - B_1^{\alpha}  \\
- E_3^{\alpha} & -B_2^{\alpha}  &  B_1^{\alpha} & 0
\ea
\right \rbrace .
\ee
This form is quite similar to that of field strength in
electrodynamics, but with an extra group index $\alpha$.
The component $E_i^{\alpha}$ of field strength is called
gravitoelectric field, and $B_i^{\alpha}$ is called
gravitomagnetic field. Traditional Newtonian gravity
is transmitted by $E_i^{\alpha}$. The $\alpha=0$ components
$B_i^0$ and $E_i^0$ correspond to the gravitoelectromagnetic
field defined in literature \cite{m05,m06,m07,m08,m09,m10}.
\\

In most cases, gravitational field is weak and its self
interactions can be neglected. So, eq.(\ref{2.19}) can
be simplified to
\be
F_{\mu\nu}^{\alpha} = \partial_{\mu} C_{\nu}^{\alpha}
-\partial_{\nu} C_{\mu}^{\alpha}.
\ee
The gravitomagnetic and gravitoelectric field have much
familiar forms
\be
B_i^{\alpha} = \partial_k C_j^{\alpha} - \partial_j C_k^{\alpha},
\ee
\be
E_i^{\alpha} = \partial_t C_i^{\alpha} - \partial_i C_0^{\alpha}
\ee
From definitions eq.(\ref{12.2}), we can prove that
\be
\nabla \cdot  \svec{B}^{\alpha} =0,
\label{12.5}
\ee
\be
\frac{\partial}{\partial t} \svec{B}^{\alpha}
+ \nabla \times \svec{E}^{\alpha} =0.
\label{12.6}
\ee
From eq.(\ref{12.5}), we know that gravitomagnetic field
is source free. It is found that gravitomagnetic field
is generated  by moving objects, and  transmit gravitomagnetic
interactions between two rotating objects.  \\

\section{Gravitational Lorentz Force}
\setcounter{equation}{0}

A particle which moves in a gravitomagnetic field will feel
a force that is perpendicular to its motion.
In electrodynamics, this force is usually
called Lorentz force. As an example, we discuss gravitational interactions
between gravitational field and Dirac field.
The lagrangian for gravitational gauge interactions of
Dirac field is\cite{18,wu06}
\be
{\cal L}_0 =
- \bar{\psi} (\gamma^{\mu} D_{\mu} + m) \psi
-  \frac{1}{4} C^{\mu\nu\rho\sigma}_{\alpha\beta}
F_{\mu \nu}^{\alpha} F_{\rho \sigma}^{\beta} ,
\label{6.2}
\ee
where
\be
C^{\mu\nu\rho\sigma}_{\alpha\beta}
= \frac{1}{4} \eta^{\mu \rho}
\eta^{\nu \sigma} g_{\alpha \beta}
+ \frac{1}{2} \eta^{\mu \rho}
G^{-1 \nu}_{\beta} G^{-1 \sigma}_{\alpha}
- \eta^{\mu \rho}
G^{-1 \nu}_{\alpha} G^{-1 \sigma}_{\beta}.
\ee
So, the interaction Lagrangian is
\be
{\cal L}_I = g  \bar\psi
\gamma^{\mu} \partial_{\alpha} \psi C_{\mu}^{\alpha}.
\label{12.14}
\ee
For Dirac field, the gravitational energy-momentum of Dirac field is
\be
T_{g \alpha}^{\mu} = \bar\psi
\gamma^{\mu} \partial_{\alpha} \psi.
\label{12.15}
\ee
Substitute eq.(\ref{12.15}) into eq.(\ref{12.14}), we get
\be
{\cal L}_I = g  T_{g \alpha}^{\mu} C_{\mu}^{\alpha}.
\label{12.16}
\ee
The interaction Hamiltonian density ${\cal H}_I$ is
\be
{\cal H}_I = - {\cal L}_I  =
- g  T_{g \alpha}^{\mu}(y,\svec{x}) C_{\mu}^{\alpha}(y).
\label{12.17}
\ee
Suppose that the moving particle is a mass point at point $\svec{x} $,
in this case
\be
 T_{g \alpha}^{\mu}(y,\svec{x}) =  T_{g \alpha}^{\mu}
\delta(\svec{y} - \svec{x} ),
\label{12.18}
\ee
where $ T_{g \alpha}^{\mu}$ is independent of space coordinates.
Then, the interaction Hamiltonian $H_I$ is
\be
H_I = \int {\rm d}^3 \svec{y} {\cal H}_I (y)
= - g \int {\rm d}^3 \svec{y}
 T_{g \alpha}^{\mu}(y,\svec{x}) C_{\mu}^{\alpha}(y).
\label{12.19}
\ee
The gravitational force that acts on the mass point is
\be
f_i = g \int {\rm d}^3y  T^{\mu}_{g \alpha}(y,\svec{x})
F^{\alpha}_{i \mu}
+g \int {\rm d}^3y  T^{\mu}_{g \alpha}(y,\svec{x})
\frac{\partial}{\partial y^{\mu}} C_i^{\alpha}.
\label{12.20}
\ee
For quasi-static system, if we omit higher order contributions, the
second term in the above relation vanish.
For a mass point, using the technique of Lorentz covariance analysis,
we can proved that
\be
P_{g \alpha} U^{\mu} = \gamma T_{g \alpha}^{\mu},
\label{12.21}
\ee
where $U^{\mu}$ is velocity, $\gamma$ is the rapidity, and
$P_{g \alpha}$ is the gravitational energy-momentum. According
eq.(\ref{12.18}), $P_{g \alpha}$ is given by
\be
P_{g \alpha} = \int {\rm d}^3 \svec{y}
 T_{g \alpha}^0(y) =  T_{g \alpha}^0.
\label{12.22}
\ee
Using all these relations and eq.(\ref{12.2}), we get
\be \label{12.22a}
\svec{f} = -g  P_{g \alpha} \svec{E}^{\alpha}
- g  P_{g \alpha} \svec{v} \times \svec{B}^{\alpha}.
\ee
For quasi-static system, the dominant contribution of the above
relation is
\be
\svec{f} = g  M \svec{E}^0
+ g  M \svec{v} \times \svec{B}^0,
\label{12.23}
\ee
where $\svec{v}= \svec{U}/\gamma$ is the velocity of the mass point.
The first term of eq.(\ref{12.23}) is the classical Newton's gravitational
interactions. The second term of eq.(\ref{12.23}) is the Lorentz force. The
direction of this force is perpendicular to the direction of the motion
of the mass point. When
the mass point is at rest or is moving along the direction
of the gravitomagnetic field, this force vanishes.
Lorentz force should have some influences for cosmology, for the rotation
of galaxy will generate gravitomagnetic field and this gravitomagnetic
field will affect the motion of stars and affect the large scale
structure of galaxy. \\

\section{Repulsive Component of Gravitational Interactions}
\setcounter{equation}{0}

The classical gravitational interactions are always attractive,
but sometimes, there are repulsive components in gravity.
The gravitational force is given by
eq.(\ref{12.22a}). The first term corresponds to classical gravitational
force. It is
\be
\begin{array}{rcl}
\svec{f}  &=& -  g   P_{g \alpha} \svec{E}^{\alpha}  \\
&&\\
&=& g  ( M_1 \svec{E}^0 - P_{gj} \svec{E}^j   ),
\end{array}
\label{12.60}
\ee
where $M_1$ is the gravitational mass of the mass point which
is moving in gravitational field. Suppose that the gravitational field
is generated by another mass point whose gravitational
energy-momentum is $Q_g^{\alpha}$ and its gravitational mass is $M_2$.
For quasi-static gravitational field, we can get
\be
\svec{E}^{\alpha} =
- \frac{g}{4 \pi r^3} Q_g^{\alpha} \svec{r}
\label{12.61}
\ee
Substitute eq.(\ref{12.61}) into eq.(\ref{12.60}), we get
\be
\svec{f} =  \frac{g^2}{4 \pi r^3}
\svec{r} ( - E_{1g} E_{2g} + \svec{P}_g \cdot \svec{Q}_g ),
\label{12.62}
\ee
where $E_{1g}$  and $E_{2g}$ are gravitational energy of two mass point.
From eq.(\ref{12.62}), we can see that, if $\svec{P}_g \cdot \svec{Q}_g$ is
positive, the corresponding component gravitational force between two momenta
is repulsive. Because $E_{1g} E_{2g} \ge \svec{P}_g \cdot \svec{Q}_g$,
the total gravitational force between two mass point is always
attractive. \\

Suppose that a Lorentz transformation along the
direction which is perpendicular to the direction of the
gravitational force $\svec{f}$, the left hand side of eq.(\ref{12.62})
is invariant. Under this transformation, the gravitational energy
of both mass point will be changed, so $E_{1g} E_{2g}$ will be changed,
but the sum $- E_{1g} E_{2g} + \svec{P}_g \cdot \svec{Q}_g$ keep
invariant. Therefore, the appearance of the component gravitational
force between two momenta satisfies the requirement of Lorentz
symmetry.  \\

\section{Discussions}
\setcounter{equation}{0}

Eq.(\ref{12.22a}) is obtained from the interaction Lagrangian, it is
deduced without concerning dynamics of gravitational gauge field. It is
known that the selection of the Lagrangian of pure gravitational
gauge field is not unique, but this ambiguity does not affect
Eq.(\ref{12.22a}). Different selection of the Lagrangian of
pure gravitational gauge field only gives out different
$\svec{E}^{\alpha}$ and $\svec{B}^{\alpha}$, it does not affect
classical Newtonian gravitational interactions and gravitational
Lorentz force.  \\

In classical Newton's theory of gravity, gravitational force between
tow objects is always along the line which connects the center of mass
of the two objects. Because of the gravitational Lorentz force, there
exists components force which is perpendicular to the line. In most cases,
this orthogonal component is  much weaker than the traditional
Newtonian gravity. \\

Gravitational Lorentz force is independent of electric charge, so it
is different from the electric Lorentz force. Only charged particle
can feel electric Lorentz force, but all particles can feel gravitational
Lorentz force. Using this property, we can discriminate gravitational
Lorentz force from electric Lorentz force, and discriminate gravitomagnetic
field from the electromagnetic magnetic field.
Effects of gravitational Lorentz force should be observable and can be
observed by astrophysical observations. It is known that there are
some experiments designed to test gravitomagnetic effects. 
\\


\begin{thebibliography}{99}

\bibitem{01} Isaac Newton, {\it Mathematical Principles of Natural Philosophy},
    (Camgridge University Press, 1934).
\bibitem{02} Albert Einstein, Annalen der Phys., {\bf 49} (1916) 769 .
\bibitem{03} Albert Einstein, Zeits. Math. und Phys. {\bf 62} (1913) 225.
\bibitem{04} S.~Weinberg, {\it Gravitation and Cosmology:
    Principles and Applications of the General Theory
    of Relativity}, (John Wiley, New York, 1972).
\bibitem{05} N.~Straumann, {\it General Relativity and
    Relativistic Astrophysics}, (Springer-Verlag, Berlin Heidelberg
    New York Tokyo, 1984).
\bibitem{wu01} Ning Wu, "Gauge Theory of Gravity", hep-th/0109145
\bibitem{wu02} Ning Wu, Commun. Theor. Phys. (Beijing, China)
        {\bf 38} (2002): 151-156.
\bibitem{wu03} Ning Wu, "Quantum Gauge Theory of Gravity",
    hep-th/0112062
\bibitem{wu04} Ning Wu, "Quantum Gauge Theory of Gravity", talk given
        at Meeting of the Devision of Particles and Fields
        of American Physical Society at the College of
        William \& Mery(DPF2002), May 24-28, 2002,
        Williamsburg, Virgia, USA; hep-th/0207254;
        Transparancy can be obtained from:
        http://dpf2002.velopers.net/talks\_pdf/33talk.pdf
\bibitem{wu05} Ning Wu, "Renormalizable Quantum Gauge General Relativity"
        gr-qc/0309041.
\bibitem{wu06} Ning Wu, Commun. Theor. Phys. (Beijing, China)
        {\bf 42} (2004): 543-552.
\bibitem{12} Ning Wu, Commun. Theor. Phys. (Beijing, China)
        {\bf 38} (2002): 322-326.
\bibitem{13} Ning Wu, Commun. Theor. Phys. (Beijing, China)
        {\bf 38} (2002): 455-460.
\bibitem{14} Ning Wu, Commun. Theor. Phys. (Beijing, China)
        {\bf 39} (2003): 561-568.
\bibitem{15} Ning Wu, Commun. Theor. Phys. (Beijing, China)
        {\bf 36}(2001) 169-172.
\bibitem{16} Ning Wu, Commun. Theor. Phys. (Beijing, China)
        {\bf 39} (2003): 671-674.
\bibitem{17} Ning Wu, Commun. Theor. Phys. (Beijing, China)
        {\bf 41}(2004) 567-572; hep-th/0307225.
\bibitem{m01} Maxwell J.C., Phil. Trans., {\bf 155},  (1865) 492.
\bibitem{m02} G. Holzmuller, Z. Math. Phys., {\bf 15}, (1870) 69.
\bibitem{m03} F. Tisserand, Compte Rendu hebdomadaire des sceances de
            L'Academie des sciences, {\bf 75} (1872) 760; {\bf 110} (1890) 313.
\bibitem{m04} O. Heaviside, Electromagnetic Theory, The Electrician Printing
            and Publishing Co., London, 1894, Vol. I, Appendix B;
            O. Heaviside, The Electrician, {\bf 31} (1893) 281-282 and 359.
\bibitem{m05} A. Einstein, Phys. Z., {\bf 14}, (1913) 1261.
\bibitem{m06} H. Thirring, Phys. Z., {\bf 19}, (1918) 204;
            {\bf 19}, (1918) 33; {\bf 22}, (1921) 29.
\bibitem{m07} J. Lense and H. Thirring, Phys. Z., {\bf 19}, (1918) 156;
            B. Mashhoon, F.W. Hehl and D.S. Thesis, Gen. Rel. Grav.,
            {\bf 16}, (1984) 711.
\bibitem{m08} M.L. Ruggiero and A. Tartaglia, gr-qc/0207065.
\bibitem{m09} B. Mashhoon, gr-qc/0311030.
\bibitem{m10} G. Schafer, gr-qc/0407116.
\bibitem{18} Ning Wu, Commun. Theor. Phys. (Beijing, China)
        {\bf 41}(2004) 381-384;





\end{thebibliography}
\end{document}